\documentclass[twocolumn,showpacs,preprintnumbers,amsmath,amssymb,prd]{revtex4}

\usepackage{graphicx}% Include figure files
\usepackage{dcolumn}% Align table columns on decimal point
\usepackage{bm}% bold math

\newcommand{\ipb}{\ensuremath{\mathrm{pb^{-1}}}}

\newcommand{\TeV}{\ensuremath{\mathrm{Te\kern -0.1em V}}}
\newcommand{\TeVc}{\ensuremath{\mathrm{Te\kern -0.1em V\!/}c}}
\newcommand{\TeVcc}{\ensuremath{\mathrm{Te\kern -0.1em V\!/}c^2}}
\newcommand{\GeV}{\ensuremath{\mathrm{Ge\kern -0.1em V}}}
\newcommand{\GeVc}{\ensuremath{\mathrm{Ge\kern -0.1em V\!/}c}}
\newcommand{\GeVcc}{\ensuremath{\mathrm{Ge\kern -0.1em V\!/}c^2}}
\newcommand{\MeV}{\ensuremath{\mathrm{Me\kern -0.1em V}}}
\newcommand{\MeVc}{\ensuremath{\mathrm{Me\kern -0.1em V\!/}c}}
\newcommand{\MeVcc}{\ensuremath{\mathrm{Me\kern -0.1em V\!/}c^2}}

\newcommand{\cdfref}[1]{\ignorespaces $^{#1}$}

\font\eightit=cmti8                                                                                

\begin{document}
%==============================================================================

\title{
%{\raggedleft\small
%\vspace*{-1.5cm}
%FERMILAB-PUB-04/xxx-E\\
%CDF/PUB/BOTTOM/CDFR/6719\\
%(Subm. Phys. Rev. D Rapid Comm.)\\
%}
%\vspace*{0.9cm}
Measurement of the Moments of the Hadronic Invariant Mass
Distribution in Semileptonic ${\boldmath B}$ Decays}% Force line breaks with \\

\author{ 
%\font\eightit=cmti8
%\hfilneg
%\begin{sloppypar}
\noindent 
D.~Acosta,\cdfref {16} J.~Adelman,\cdfref {12} T.~Affolder,\cdfref 9 T.~Akimoto,\cdfref {54}
M.G.~Albrow,\cdfref {15} D.~Ambrose,\cdfref {43} S.~Amerio,\cdfref {42}  
D.~Amidei,\cdfref {33} A.~Anastassov,\cdfref {50} K.~Anikeev,\cdfref {15}
% ...
%\end{sloppypar}
\vskip .026in
%\begin{center}
(CDF Collaboration)
%\end{center}
\vskip .026in
%\begin{center}
\cdfref 1  {\eightit Institute of Physics, Academia Sinica, Taipei, Taiwan 11529, 
Republic of China} \\
\cdfref 2  {\eightit Argonne National Laboratory, Argonne, Illinois 60439} \\
\cdfref 3  {\eightit Institut de Fisica d'Altes Energies, Universitat Autonoma
de Barcelona, E-08193, Bellaterra (Barcelona), Spain} \\
\cdfref 4  {\eightit Istituto Nazionale di Fisica Nucleare, University of Bologna,
I-40127 Bologna, Italy} \\
\cdfref 5  {\eightit Brandeis University, Waltham, Massachusetts 02254} \\
\cdfref 6  {\eightit University of California at Davis, Davis, California  95616} \\
\cdfref 7  {\eightit University of California at Los Angeles, Los 
Angeles, California  90024} \\
\cdfref 8  {\eightit University of California at San Diego, La Jolla, California  92093} \\ 
\cdfref 9  {\eightit University of California at Santa Barbara, Santa Barbara, California 
93106} \\ 
\cdfref {10} {\eightit Instituto de Fisica de Cantabria, CSIC-University of Cantabria, 
39005 Santander, Spain} \\
\cdfref {11} {\eightit Carnegie Mellon University, Pittsburgh, PA  15213} \\
\cdfref {12} {\eightit Enrico Fermi Institute, University of Chicago, Chicago, 
Illinois 60637} \\
\cdfref {13}  {\eightit Joint Institute for Nuclear Research, RU-141980 Dubna, Russia}
\\
\cdfref {14} {\eightit Duke University, Durham, North Carolina  27708} \\
\cdfref {15} {\eightit Fermi National Accelerator Laboratory, Batavia, Illinois 
60510} \\
\cdfref {16} {\eightit University of Florida, Gainesville, Florida  32611} \\
\cdfref {17} {\eightit Laboratori Nazionali di Frascati, Istituto Nazionale di Fisica
               Nucleare, I-00044 Frascati, Italy} \\
\cdfref {18} {\eightit University of Geneva, CH-1211 Geneva 4, Switzerland} \\
\cdfref {19} {\eightit Glasgow University, Glasgow G12 8QQ, United Kingdom}\\
\cdfref {20} {\eightit Harvard University, Cambridge, Massachusetts 02138} \\
\cdfref {21} {\eightit The Helsinki Group: Helsinki Institute of Physics; and Division of
High Energy Physics, Department of Physical Sciences, University of Helsinki, FIN-00044, Helsinki, Finland}\\
\cdfref {22} {\eightit Hiroshima University, Higashi-Hiroshima 724, Japan} \\
\cdfref {23} {\eightit University of Illinois, Urbana, Illinois 61801} \\
\cdfref {24} {\eightit The Johns Hopkins University, Baltimore, Maryland 21218} \\
\cdfref {25} {\eightit Institut f\"{u}r Experimentelle Kernphysik, 
Universit\"{a}t Karlsruhe, 76128 Karlsruhe, Germany} \\
\cdfref {26} {\eightit High Energy Accelerator Research Organization (KEK), Tsukuba, 
Ibaraki 305, Japan} \\
\cdfref {27} {\eightit Center for High Energy Physics: Kyungpook National
University, Taegu 702-701; Seoul National University, Seoul 151-742; and
SungKyunKwan University, Suwon 440-746; Korea} \\
\cdfref {28} {\eightit Ernest Orlando Lawrence Berkeley National Laboratory, 
Berkeley, California 94720} \\
\cdfref {29} {\eightit University of Liverpool, Liverpool L69 7ZE, United Kingdom} \\
\cdfref {30} {\eightit University College London, London WC1E 6BT, United Kingdom} \\
\cdfref {31} {\eightit Massachusetts Institute of Technology, Cambridge,
Massachusetts  02139} \\   
\cdfref {32} {\eightit Institute of Particle Physics: McGill University,
Montr\'{e}al, Canada H3A~2T8; and University of Toronto, Toronto, Canada
M5S~1A7} \\
\cdfref {33} {\eightit University of Michigan, Ann Arbor, Michigan 48109} \\
\cdfref {34} {\eightit Michigan State University, East Lansing, Michigan  48824} \\
\cdfref {35} {\eightit Institution for Theoretical and Experimental Physics, ITEP,
Moscow 117259, Russia} \\
\cdfref {36} {\eightit University of New Mexico, Albuquerque, New Mexico 87131} \\
\cdfref {37} {\eightit Northwestern University, Evanston, Illinois  60208} \\
\cdfref {38} {\eightit The Ohio State University, Columbus, Ohio  43210} \\  
\cdfref {39} {\eightit Okayama University, Okayama 700-8530, Japan}\\  
\cdfref {40} {\eightit Osaka City University, Osaka 588, Japan} \\
\cdfref {41} {\eightit University of Oxford, Oxford OX1 3RH, United Kingdom} \\
\cdfref {42} {\eightit University of Padova, Istituto Nazionale di Fisica 
          Nucleare, Sezione di Padova-Trento, I-35131 Padova, Italy} \\
\cdfref {43} {\eightit University of Pennsylvania, Philadelphia, 
        Pennsylvania 19104} \\   
\cdfref {44} {\eightit Istituto Nazionale di Fisica Nucleare, University and Scuola
               Normale Superiore of Pisa, I-56100 Pisa, Italy} \\
\cdfref {45} {\eightit University of Pittsburgh, Pittsburgh, Pennsylvania 15260} \\
\cdfref {46} {\eightit Purdue University, West Lafayette, Indiana 47907} \\
\cdfref {47} {\eightit University of Rochester, Rochester, New York 14627} \\
\cdfref {48} {\eightit The Rockefeller University, New York, New York 10021} \\
\cdfref {49} {\eightit Istituto Nazionale di Fisica Nucleare, Sezione di Roma 1,
University di Roma ``La Sapienza," I-00185 Roma, Italy}\\
\cdfref {50} {\eightit Rutgers University, Piscataway, New Jersey 08855} \\
\cdfref {51} {\eightit Texas A\&M University, College Station, Texas 77843} \\
\cdfref {52} {\eightit Texas Tech University, Lubbock, Texas 79409} \\
\cdfref {53} {\eightit Istituto Nazionale di Fisica Nucleare, University of Trieste/\
Udine, Italy} \\
\cdfref {54} {\eightit University of Tsukuba, Tsukuba, Ibaraki 305, Japan} \\
\cdfref {55} {\eightit Tufts University, Medford, Massachusetts 02155} \\
\cdfref {56} {\eightit Waseda University, Tokyo 169, Japan} \\
\cdfref {57} {\eightit Wayne State University, Detroit, Michigan  48201} \\
\cdfref {58} {\eightit University of Wisconsin, Madison, Wisconsin 53706} \\
\cdfref {59} {\eightit Yale University, New Haven, Connecticut 06520}
%\end{center}
}
\noaffiliation
\date{\today}

\begin{abstract}
\noindent
Using 180 $\rm{pb}^{-1}$ of data collected with the CDF II detector
at the Tevatron, we measure the first two moments of the hadronic
invariant mass-squared distribution in charmed semileptonic $B$ decays.  From these
we determine the non-perturbative Heavy Quark Effective Theory parameters $\Lambda$ and
$\lambda_1$ used to relate the $B$ meson semileptonic branching ratio
to the CKM matrix element $|V_{cb}|$. For a minimum lepton momentum of 0.7~GeV/$c$ in the $B$
rest frame we measure the first two moments of the ${D}^{**} \to
{D}^{(*)}\pi$ component to be $\langle m^2_{D^{**}}\rangle = (5.83 \pm 0.16_{\mathrm{stat}} 
\pm 0.08_{\mathrm{syst}})\ \mathrm{GeV}^2/c^4$ and
$\langle(m^2_{D^{**}} - \langle m^2_{D^{**}}\rangle )^2\rangle =
(1.30 \pm 0.69_{\mathrm{stat}} \pm 0.22_{\mathrm{syst}})\ {\mathrm{GeV^4}}/c^8$.  Combining these
with the discrete mass terms from the $D$ and ${D}^*$ mesons, we find the
total 
moments to be $\langle M_{X_c}^2\rangle-\overline{m}_D^2 
= (0.467 \pm 0.038_{\mathrm{stat}} \pm 0.068_{\mathrm{syst}})\ {\mathrm{GeV^2}}/c^4$ and
$\langle(M_{X_c}^2 - \langle M_{X_c}^2\rangle )^2\rangle
= (1.05  \pm 0.26_{\mathrm{stat}} \pm 0.13_{\mathrm{syst}})\ {\mathrm{GeV^4}}/c^8$, where $\overline{m}_D$
is the spin-averaged $D$ mass.  The systematic error is dominated by
the uncertainties in the world-average branching ratios used to combine
the $D$, ${D}^*$, and ${D}^{**}$ contributions. The analysis makes
no assumptions about the shape or resonant structure of the ${D}^{**} \to
{D}^{(*)}\pi$ invariant mass distribution.
%-------------------------------------------------------------------------------
\end{abstract}
\maketitle
%===============================================================================
\section{Introduction}
\label{sec:intro}
In order to constrain the length of the side opposite the angle $\beta$ in the
Cabibbo-Kobayashi-Maskawa (CKM) unitarity triangle, a precise measurement of the ratio
$|V_{ub}|/|V_{cb}|$ is needed. The matrix element
$|V_{cb}|$ is generally extracted from semileptonic $B$ decays.
Currently, the most precise method is based on the measurement of the inclusive
semileptonic partial width into charm, $\Gamma_{sl}=\Gamma(B\to X_cl\nu_l)$.
The Operator Product Expansion (OPE) applied to Heavy Quark Effective Theory (HQET) relates the 
experimental determination of $\Gamma_{sl}$ to $|V_{cb}|$~\cite{ref:falk-gremm,ref:bauer}. The relationship
takes the form of an expansion in inverse powers of the $B$ mass, $m_B$. At
each order in the expansion, new free non-perturbative
parameters
enter: one (${\Lambda}$) at order $1/m_B$, two 
($\lambda_1$ and $\lambda_2$) at order $1/m_B^2$, six at order $1/m_B^3$, etc.
In order to extract $|V_{cb}|$ from $\Gamma_{sl}$ some external information on
these parameters is needed.

The same theoretical framework that predicts the value of $\Gamma_{sl}$
predicts the value of any weighted integral of the differential rate $d\Gamma_{sl}/ds_H$, provided the weight is a
smooth function of $s_H\equiv M_{X_c}^2$. Using weight functions 
$(s_H-\overline{m}_D^2)$ and
$(s_H-\langle s_H\rangle)^2$, 
with $\overline{m}_D = 0.25m_D+0.75m_{D^*}$,
one can
define the first two moments of the hadronic mass distribution:
\begin{eqnarray}
M_1
 &=& \int_{s_H^{min}}^{s_H^{max}} ds_H \,\left(s_H-\overline{m}_D^2\right)\,
\frac{1}{\Gamma_{sl}}\frac{d\Gamma_{sl}}{ds_H} \ ,
\nonumber \\
M_2
&=& \int_{s_H^{min}}^{s_H^{max}} ds_H \,\left(s_H-\langle s_H\rangle\right)^2 \,
\frac{1}{\Gamma_{sl}}\frac{d\Gamma_{sl}}{ds_H} \ ,
\label{eq:M1M2}
\end{eqnarray}
which are simply the shifted mean and variance of the $M_{X_c}^2$ 
distribution in
semileptonic charmed decays of $B$ mesons. The
moments are not sensitive to $|V_{cb}|$,
but they are more sensitive to the non-perturbative parameters of HQET than
$\Gamma_{sl}$ itself is. Therefore, measuring the moments provides a useful
constraint on the HQET parameters which improves the 
overall precision on $|V_{cb}|$ as determined from $\Gamma_{sl}$. This is the 
purpose of this analysis. Since $\lambda_2$ is well determined from the values of
the hyperfine mass splittings in the $B$ and $D$ meson systems~\cite{ref:bauer}, only
$\Lambda$ and $\lambda_1$ are studied here.

The $s_H$ distribution in $B^-\to X_c^0 l^- \overline{\nu}_l$ decays can be split
into three contributions corresponding to $X_c^0=D^0,D^{*0},D^{**0}$. Here $D^{**0}$ stands for any neutral
charmed state, resonant or not, other than $D^0$, $D^{*0}$. The differential mass-squared spectrum can be written as:
\begin{eqnarray}
\frac{1}{\Gamma_{sl}}\frac{d\Gamma_{sl}}{ds_H} &=& \frac{\Gamma_0}{\Gamma_{sl}}\cdot\delta(s_H-m_{D^0}^2)
+ \frac{\Gamma_*}{\Gamma_{sl}}\cdot\delta(s_H-m_{D^{*0}}^2) \nonumber \\
&+& \left(1-\frac{\Gamma_0}{\Gamma_{sl}}-\frac{\Gamma_*}{\Gamma_{sl}}\right)\cdot f^{**}(s_H)\ ,
\label{eq:strategy}
\end{eqnarray}
where $\Gamma_{sl}$ is now the inclusive $B^-$ semileptonic width, $\Gamma_0$ and $\Gamma_*$ are the exclusive
$B^-$ partial
widths to $D^0l^-\overline{\nu}_l$ and $D^{*0}l^-\overline{\nu}_l$ respectively, and $f^{**}(s_H)$ is the normalized
hadronic invariant mass-squared distribution in the $D^{**0}$ channel. We use world-average values of
$\Gamma_{0}/\Gamma_{sl}$, $\Gamma_{*}/\Gamma_{sl}$, $m_{D^0}$ and $m_{D^{*0}}$ from the 
Particle Data Group~\cite{ref:pdg04} and concentrate on 
measuring $f^{**}(s_H)$. In this way, we have only to measure the invariant mass distribution for the $D^{**0}$
component without having to determine the $D^0$, $D^{*0}$ components or the relative normalizations between those
and the $D^{**0}$ channel. 
The $D^{**0}$ spectrum is not well known, and includes, at least, two narrow and two wide states, together with a
possible non-resonant $D^{(*)}{\mathrm n}\pi$ contribution. The measurement of the $D^{**0}$ spectrum is the main task of this analysis. 
We assume that the $D^{(*)}\pi l^- \overline{\nu}_l$  decays of $B^-$ saturate the difference between its inclusive 
semileptonic decay rate and the sum of its exclusive decay rates to $D^0l^- \overline{\nu}_l$ and $D^{*0}l^- \overline{\nu}_l$.
We neglect all modes with additional pions in the final state, as well as $D^{**0}\to D_s^{(*)+}K^-$.

Only $D^{(*)+}\pi^-$ decays (charge conjugated channels are implicitly included throughout the paper)
are reconstructed. 
Contributions to the $s_H$ distribution from decays with neutral particles are included by applying isospin factors
to the charged modes.
Feed-down from one channel to another due to unmeasured neutral particles is subtracted statistically using the data
themselves and isospin relations, as explained in Section~\ref{sec:back}.
\section{Data Analysis}
\label{sec:data}
The analysis uses a data sample of $\overline{p}p$ collisions at $\sqrt{s}=1.96~\TeV$
with an integrated luminosity of about $180~\ipb$, collected between February 2002 and
August 2003 with the upgraded Collider Detector
at the Fermilab Tevatron (CDF~II). A description of the detector can be found in~\cite{ref:detector}.
The relevant components for this analysis include a tracking
system composed of a silicon strip vertex detector (SVX~II)
surrounded by an open cell drift chamber system 
(COT).
The SVX~II detector comprises five concentric layers of double-sided sensors
located at radii between 2.5 and 10.6 cm,
while the COT provides 96 measurements (including axial and stereo) out to a radius of 132~cm.
The  central tracking
system is immersed in a 1.4~T solenoidal magnetic field.
Two sampling calorimeters surround the magnetic coil.
A set of proportional chambers inside the electromagnetic calorimeter
provides information on the shower profile for use in
electron identification. Muon candidates are identified with two
sets of multi-layer drift chambers, one located
outside the calorimeters and the other behind an additional 60~cm-thick iron shield.

Decays $B\to D^{(*)+} \pi^- l^- X$, where $l$ stands for electron or muon, were recorded using a trigger
that requires a lepton $l$ and a track displaced from the interaction point~\cite{ref:SVT}. 
The lepton and the displaced track
must have transverse
momentum $p_T$ in excess of 4~GeV/$c$ and 2~GeV/$c$ respectively. 
The displaced track's impact parameter with respect to the beamline has to exceed
120~$\mu$m and be below 1~mm.
Events which pass the trigger are recorded for further analysis.
Only well-reconstructed tracks with $p_T \geq 0.4~\GeVc$ are retained.
Track parameters are corrected for
the ionization energy loss
appropriate to the mass hypothesis under consideration. 
A Monte Carlo sample of $B\to D^{**} l \nu_l$ events based on the ISGW2~\cite{ref:ISGW2} and 
Goity-Roberts~\cite{ref:Goity-Roberts} matrix elements and including a detailed simulation of the CDF~II detector
based on the GEANT~\cite{ref:GEANT} package has been used throughout the analysis. In accordance with our
assumption, only $D^{**}\to D^{(*)}\pi$ decays are generated.

Events with $D^{*+}(\to D^0 \pi_*^+)l^-$ and $D^+l^-$ combinations
are reconstructed in the decay channels $D^0\to
K^-\pi^+, \ K^-\pi^+\pi^-\pi^+,\ K^-\pi^+\pi^0$ and $D^+\to K^-\pi^+\pi^+$. Tracks with the appropriate charge 
combination are required to be consistent with a common vertex in three dimensions.
One of the tracks in the vertex 
must fulfill
the displaced trigger requirements. 
Suitable ranges are selected in the $D^0$ (1.84--1.89~GeV/$c^2$) and $D^+$ 
(1.84--1.89~GeV/$c^2$)
mass distributions.
For the $D^{*+}$ channel, an additional charged
track ($\pi_*^+$) is required, such that the $M(D^0\pi_*^+)-M(D^0)$ mass difference lies between 
0.142 and 0.147~GeV/$c^2$.
The $D^0\to K^-\pi^+\pi^0$ channel is
reconstructed from the satellite peak in the $K^-\pi^+$ mass distribution (1.50--1.70~GeV/$c^2$). In this case the
$M(K^-\pi^+\pi_*^+)-M(K^-\pi^+)$ mass difference is required to be between 0.142 and 0.155~GeV/$c^2$.
Duplicate removal is performed in the $D^0\to K^-\pi^+\pi^-\pi^+$ channel: when two
$D^{*+}$ candidates share all five 
tracks, differing only in
the kaon mass assignment, the candidate with the 
$K^-\pi^+\pi^-\pi^+$ mass closer to the nominal
$D^0$ mass is retained. No attempt has been made to further identify kaons and pions.

\begin{figure*}[ht]
\includegraphics[scale=0.38]{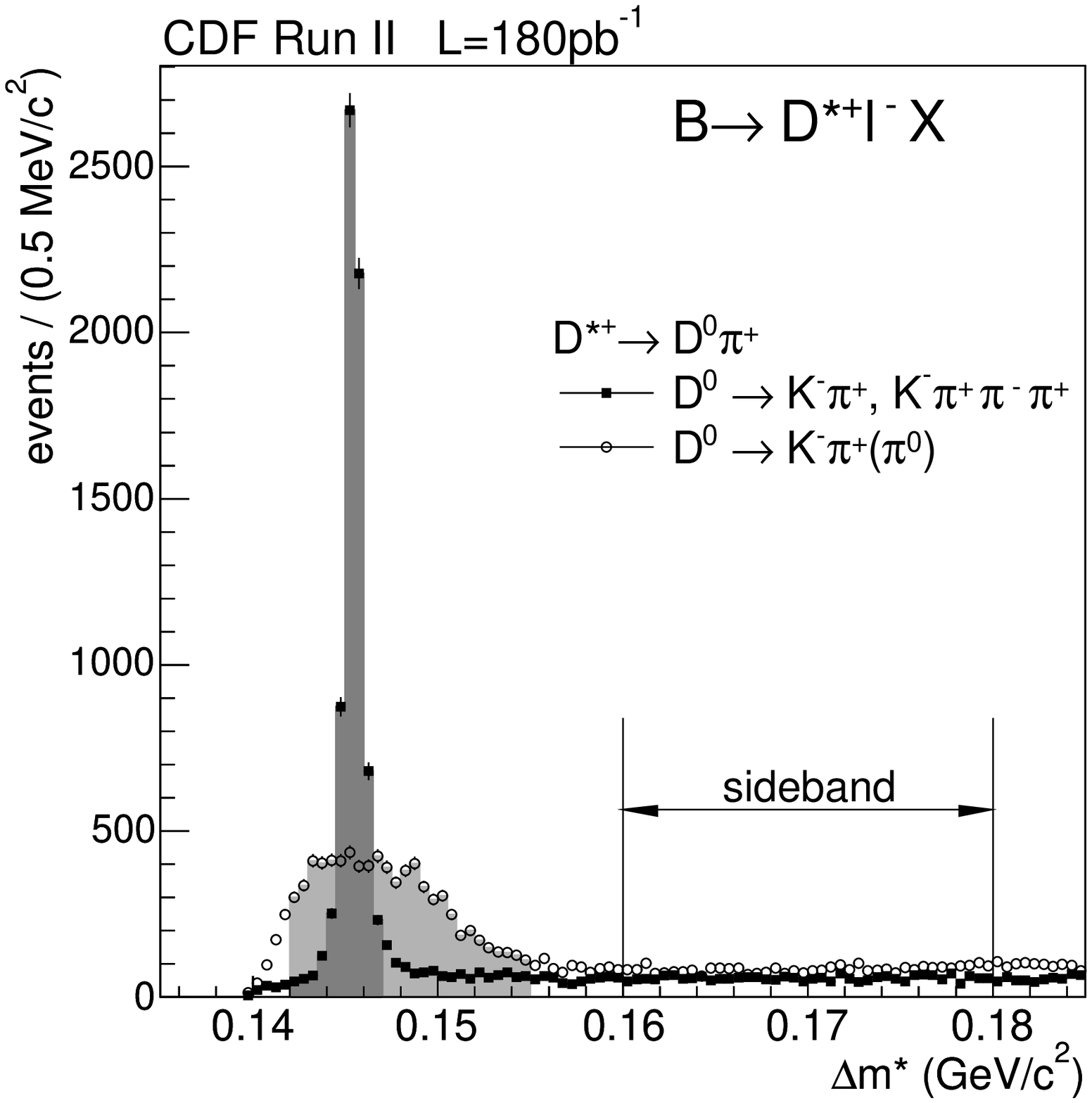}
\hspace{1cm}
\includegraphics[scale=0.38]{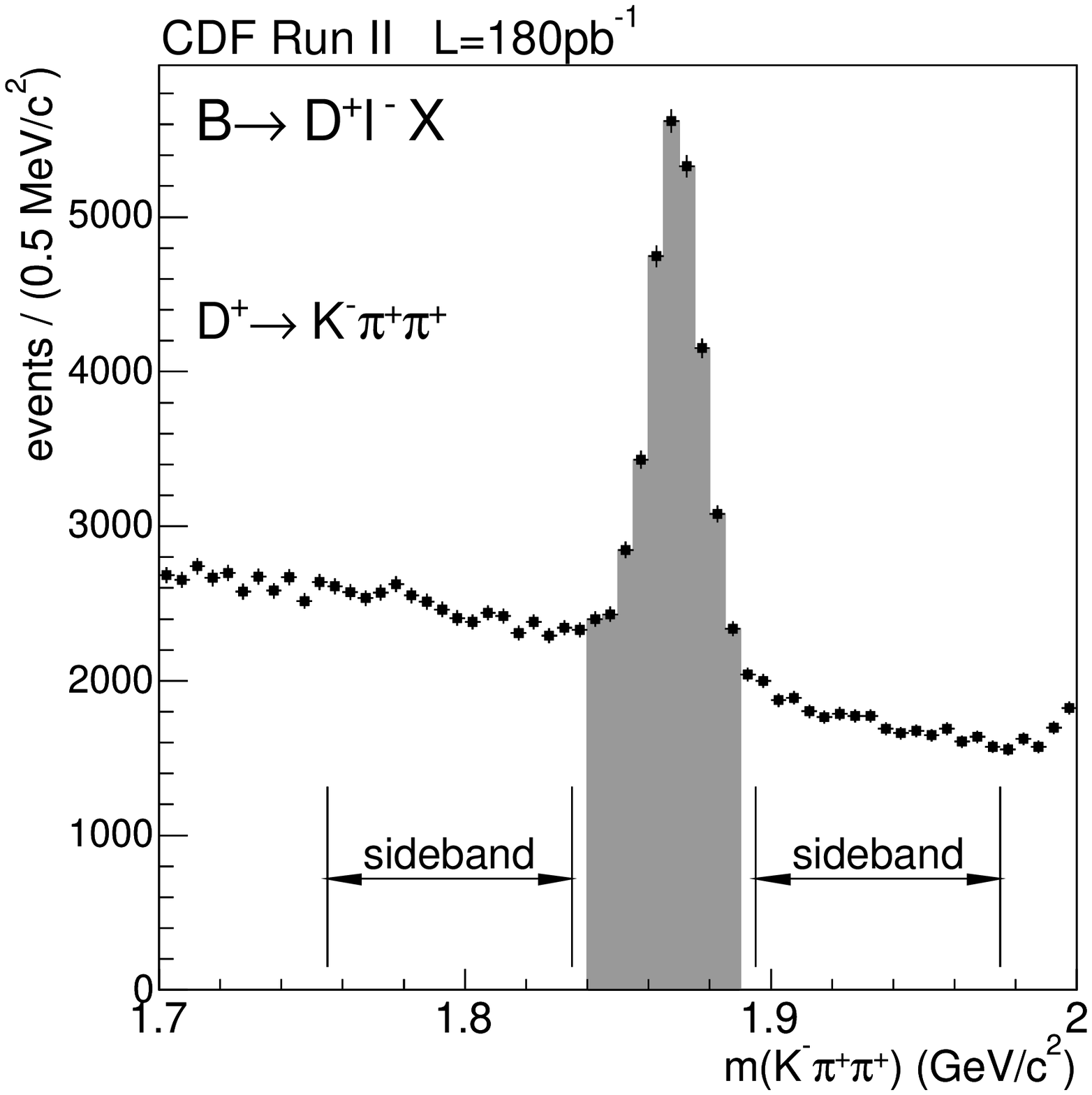}
\caption{Left: Mass difference $\Delta m^* = M(D^0\pi_*^+)-M(D^0)$ in the $D^0\to K^-\pi^+$ 
and $D^0\to K^-\pi^+\pi^+\pi^-$ channels (narrow peak) and $\Delta m^* = M(K^-\pi^+\pi_*^+)-M(K^-\pi^+)$ in the
$D^0\to K^-\pi^+\pi^0$ channel (broad peak). %in semileptonic $B$ decays. 
Right: 
Mass distribution for the $D^+\to K^-\pi^+\pi^+$ channel.
The signal areas are shaded; sideband regions are also indicated.
\label{fig:K1pK3pSatDeltaM}}
\end{figure*}

After the selection of $D^{*+}l^-$ and $D^+l^-$ combinations, we obtain $3890\pm63$, $2994\pm57$, $6638\pm98$ and
$14416\pm202$ signal
events in the $K^-\pi^+$, $K^-\pi^+\pi^-\pi^+$, $K^-\pi^+\pi^0$ and $K^-\pi^+\pi^+$ channels, respectively. 
Combinatorial backgrounds, estimated from sidebands of the $M(D^+)$ and $M(D^0\pi_*^+)-M(D^0)$
distributions, have been subtracted.
The quoted yield in the $D^+$ channel has been rescaled by a factor 0.96 to account for the 
background from $D^+_s \to K^+ K^- \pi^+$ decays where the $K^+$ is assigned the pion mass. 
Figure~\ref{fig:K1pK3pSatDeltaM} shows on the left the
$M(D^0\pi_*^+)-M(D^0)$ distributions for $D^0$ decaying into either $K^-\pi^+$ or $K^-\pi^+\pi^-\pi^+$,
and for $D^0 \to K^-\pi^+\pi^0$ , while 
the $D^+\to K^-\pi^+\pi^+$ mass distribution is plotted on the right.

\begin{figure*}[ht]
\includegraphics[scale=0.38]{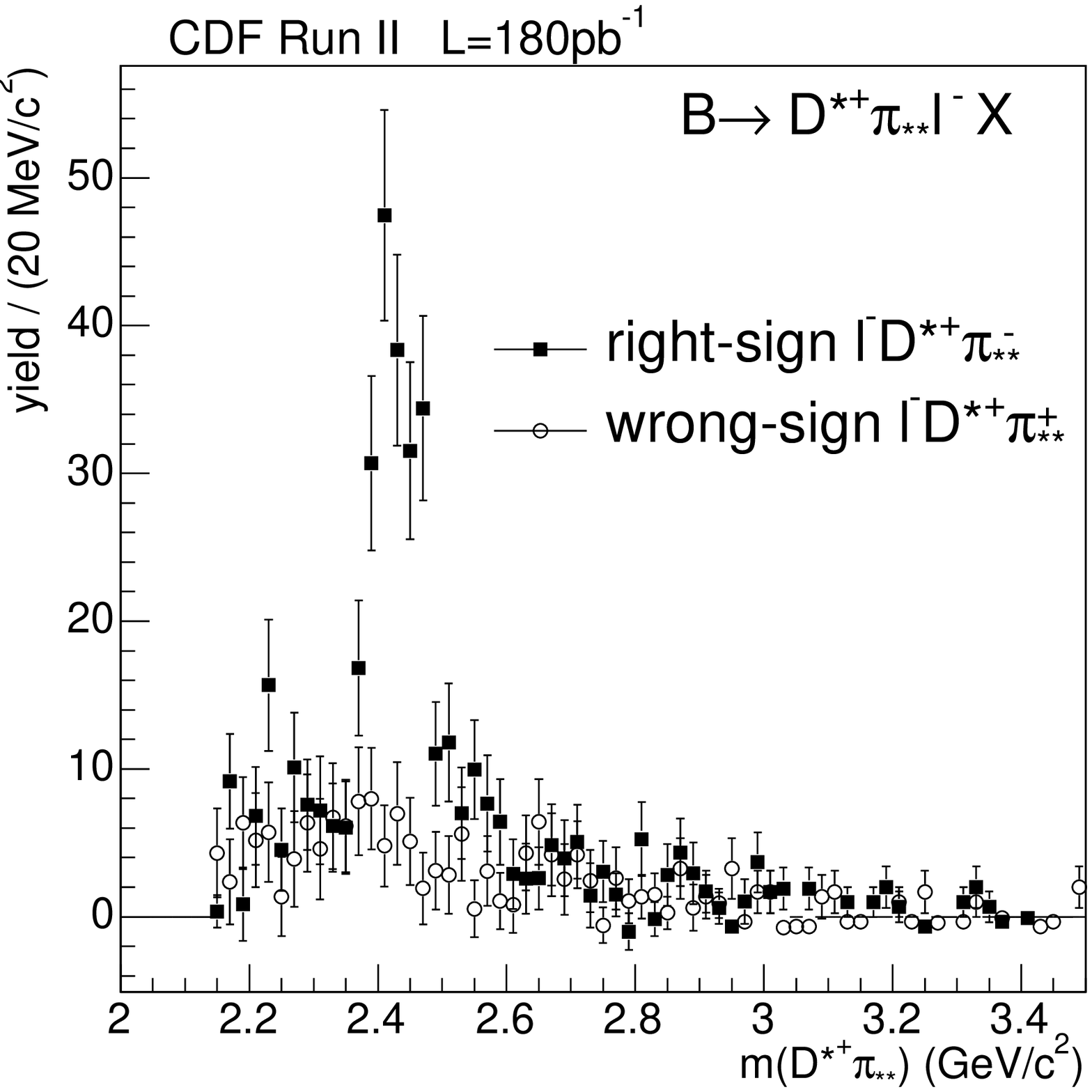}
\hspace{1cm}
\includegraphics[scale=0.38]{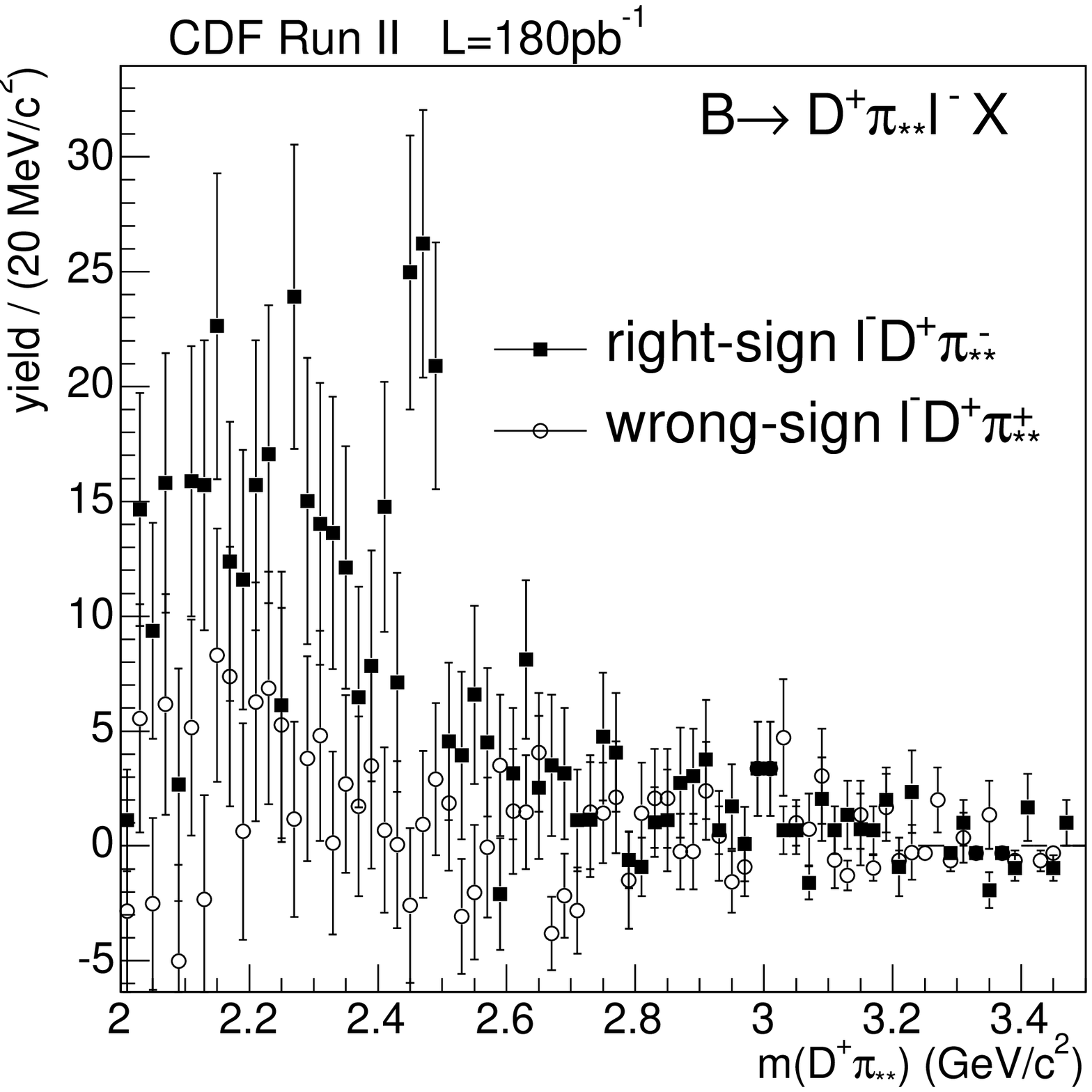}
\caption{Sideband-subtracted invariant mass distribution
for the $D^{*+}\pi_{**}$ channels (left) and for the $D^+\pi_{**}$ channel (right).
The mass regions shown are limited at 3.5 GeV$/c^2$ for illustration only. No explicit mass cut is applied in the analysis.
\label{fig:dstarplus_mass_zoom}}
\end{figure*}

The $D^{(*)+}l^-$ vertex 
(the $B$ vertex) is reconstructed in three dimensions and required to be at least 500~$\mu$m away from the beam line.
An additional pion ($\pi_{**}^-$) is then added to create full $D^{+(*)}\pi^- l^-$ candidates.
The $\pi_{**}^-$'s trajectory is required to be
at most 2.5 standard deviations away from the $B$ vertex, and at least three standard deviations away
from the beam line.
These cuts were optimized using the
$B\to D^{**} l \nu_l$ Monte Carlo 
for the signal,
and wrong-sign $\pi_{**}^+l^-$ combinations in data for the background.
The measured mass distributions in the $D^{*+}\pi_{**}$ and $D^+\pi_{**}$ channels are shown
in Fig.~\ref{fig:dstarplus_mass_zoom}.

\section{Background and Efficiency Corrections}
\label{sec:back}
The most important background sources are combinatorial background under the $D^+$ and $D^*$ mass peaks,
prompt tracks (from fragmentation or the underlying event) that fake $\pi_{**}$ candidates, and 
feed-down from $D^{*+}\to D^+\pi^0$ decays into the $D^+$ channel.
Data side-bands are used to assess combinatorial background
under the $D^+$ mass and $D^{*+}$--$D^0$ mass-difference 
peaks, and wrong-sign $\pi_{**}^+l^-$ combinations in data to characterize
the prompt background to the $\pi_{**}$
candidates. The wrong-sign pion-lepton
sample is subtracted from the right-sign sample, after performing side-band subtraction
in both. A possible difference between the rate of prompt background
with the right and the wrong $\pi_{**}$ charge  
has been studied with a sample of fully reconstructed $B$ decays 
($B^-\to J/\Psi K^-,\, B^0\to D^{(*)-}\pi^+, \, B^-\to D^0 \pi^-$)
and found to be at most 4\%.
This has been included in the systematic error. 
The bias in the
background subtraction introduced by using the same wrong-sign sample for both the optimization of the
selection and the final background subtraction has been studied 
using bootstrap~\cite{ref:boot} copies of the data
and
found to be smaller than 15\% of the statistical error on the moments.
This upper bound on the bias has been added as an additional systematical error.

Since we do not reconstruct neutral particles, events 
with a decay $B^-\to D^{*+}\pi^-_{**}l^-\overline{\nu}_l$ with the $D^{*+}$ decaying into
$D^+\pi^0$ constitute an irreducible background to the signal channel $B^-\to D^+\pi^-_{**}l^-\overline{\nu}_l$.
Using isospin symmetry, the background rate and $D^+\pi^-_{**}$ invariant mass
can be obtained from the $D^0\pi^-_{**}$ invariant
mass in $B^-\to D^{*+}\pi^-_{**}l^-\overline{\nu}_l$ decays with $D^{*+}\to D^0\pi^+_*$, which we fully 
reconstruct, after correcting for the relative efficiency using Monte Carlo. 
A small 
physics background (around 1\% in rate), coming mostly from $B\to D^{(*)}_s D^{(*)}$ decays with the 
$D^{(*)}_s$ decaying semileptonically, is subtracted using Monte Carlo predictions. 
Background from tracks faking a lepton has been studied by looking at 
wrong-sign $D^{(*)+}l^+$
combinations. It has been found that the background subtraction procedure outlined above effectively removes
any such background. Finally, the background from $B\to D^{(*)}\pi\tau\nu_\tau$ decays has been
studied with Monte Carlo and found to have a negligible effect on the determination of the moments.

Since only the shape of the mass-squared distribution for the $D^{**}$ component ($f^{**}(s_H)$ in Eq.~(\ref{eq:strategy}))
is being measured, the relevant
efficiency corrections are those that can bias the mass-squared distribution, along with
the relative efficiency for the
$D^{*+}$ and $D^+$ components of the $D^{**}$ piece. Both efficiencies are obtained, 
as a function of the mass $M(D^{**})$, from the Monte Carlo simulation.
We have checked the Monte Carlo relative efficiency predictions as a function of the $\pi_{**}$ transverse momentum
by applying the $\pi_{**}$ selection cuts to decay tracks from $D$ and $D^*$ mesons.
The efficiency variation in the data agrees well with the simulation.
Where small differences have
been found, corrections have been derived from these Monte Carlo--data comparisons. 

To compare the moments with theoretical
predictions, they must be measured with a well defined cut on $p_l^*$, the lepton momentum in the $B$
rest frame. Since we do not attempt to measure the boost of the $B$,
we cannot access $p_l^*$ directly in data. 
Instead, 
acceptance corrections are derived from Monte Carlo that turn our gradual trigger turn-on as a function of $p_l^*$
into a sharp threshold at $p_l^* = 0.7$~GeV/$c$, thereby 
correcting our measurement of the moments to a cut $p_l^* > 0.7$~GeV/$c$.
The value $0.7$~GeV/$c$ was chosen in order to minimize the correction. 
Because of the negative correlation between lepton momentum in the $B$ rest frame and $D^{**}$ mass,
the correction itself can depend on the detailed
$D^{**}$ mass spectrum in Monte Carlo. In order to assess the possible systematic error, the default $B$ decay model
has been compared to
a na\"{\i}ve phase-space $B$ semileptonic decay model. The differences in the ensuing correction
factors as a function of $M(D^{**})$ are considered as systematic errors.
\section{Results}
\label{sec:results}
\begin{figure}[htb]
\includegraphics[scale=0.46]{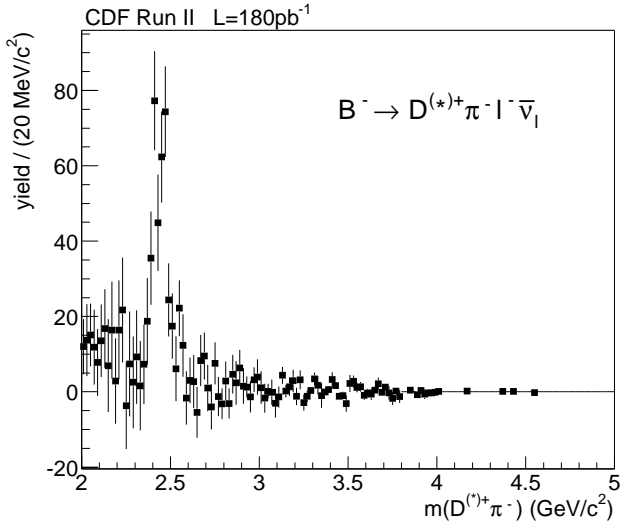}
\caption{Fully corrected invariant mass distribution $m(D^{(*)+}\pi^-)$.
The number of events in each bin has been 
background subtracted and corrected for mass-dependent and $D^{*}/D^+$ relative efficiency corrections. 
The plotted errors take into account all corrections and subtractions.\label{fig:mcorrected}
}
\end{figure}

The $D^{(*)+}\pi^-$ mass distribution is shown in Fig.~\ref{fig:mcorrected} after background subtraction and efficiency
and acceptance corrections.
The first and second moments of the $D^{**}$ component of the
mass-squared distribution, $m_1$ and $m_2$, are determined 
by simply computing the mean and variance of the distribution shown in Fig.~\ref{fig:mcorrected},
without any assumption about the shape or rate of its several components:
\begin{eqnarray}
m_1 &\equiv& \langle m_{D^{**}}^2\rangle = 
(5.83 \pm 0.16_{\mathrm{stat}} \pm 0.08_{\mathrm{syst}}) \ {\mathrm{GeV^2}}/c^4 \nonumber \\
m_2 &\equiv& \langle(m_{D^{**}}^2 - \langle m_{D^{**}}^2\rangle)^2\rangle = \nonumber \\
&& \ \ \ \ \ \ \ \ \ \ \ \ \ \, (1.30 \pm 0.69_{\mathrm{stat}} \pm 0.22_{\mathrm{syst}}) \ {\mathrm{GeV^4}}/c^8 \ , \nonumber
\end{eqnarray}
with a 61\% positive correlation.
The full moments of the hadronic mass-squared distribution, $M_1$ and $M_2$, are determined
by combining $m_1$ and $m_2$ with the $D$ and $D^*$ pieces, obtained from world-average values~\cite{ref:pdg04}:
\begin{eqnarray}
M_1\!\!&&\!\!\equiv \langle M_{X_c}^2\rangle-\overline{m}_D^2 = \nonumber \\
&&(0.467 \pm 0.038_{\mathrm{stat}} \pm 0.019_{\mathrm{exp}} \pm 0.065_{\mathrm{BR}}) 
\ {\mathrm{GeV^2}}/c^4  \nonumber \\
M_2\!\!&&\!\!\equiv \langle(M_{X_c}^2 - \langle M_{X_c}^2\rangle)^2\rangle = \nonumber \\
&&(1.05  \pm 0.26_{\mathrm{stat}} \pm 0.08_{\mathrm{exp}} \pm 0.10_{\mathrm{BR}}) 
\ {\mathrm{GeV^4}}/c^8 \ , \nonumber
\end{eqnarray}
with a 69\% positive correlation between $M_1$ and $M_2$.
Here ``BR'' refers to the uncertainty coming from the branching ratios needed for the combination of the $D$,
$D^*$ and $D^{**}$ pieces. 
For the exclusive branching ratios to $D$ and $D^*$, all available information~\cite{ref:pdg04} coming from charged and
neutral $B$ decays has been combined %, after correcting for the $p_l^*$ cut at 0.7~GeV/$c$,
using isospin invariance, leading to $\Gamma_{0}/\Gamma_{sl}= 0.203\pm 0.015$ and 
$\Gamma_{*}/\Gamma_{sl}= 0.550 \pm 0.026$ with about $30\%$ positive correlation.
The isospin-related partial widths (not the branching ratios) of $B^+$ and $B^0$ are assumed to be
identical.

Finally, using the predictions in~\cite{ref:bauer}, the HQET parameters $\Lambda$ and $\lambda_1$ are determined.
After applying constraints on the other HQET parameters coming from the known
$B$ and $D$ hyperfine mass splittings we find, in the pole scheme:
$\overline{\Lambda} \equiv \Lambda^{\mathrm{pole}} = 
( 0.397 \pm 0.078_{\mathrm{stat}} \pm 0.027_{\mathrm{exp}} \pm 0.064_{\mathrm{BR}} \pm 
                     0.058_{\mathrm{theo}})$~GeV and
$\lambda_1^{\mathrm{pole}} = 
(-0.184 \pm 0.057_{\mathrm{stat}} \pm 0.017_{\mathrm{exp}} \pm 0.022_{\mathrm{BR}} \pm
                     0.077_{\mathrm{theo}})$~GeV$^2$,
with a 79\% negative correlation. Similarly, we extract the equivalent HQET parameters in the 1S scheme:
$m_b^{1S} \equiv M_\Upsilon/2 - \Lambda^{1S} = 
( 4.654 \pm 0.078_{\mathrm{stat}} \pm 0.027_{\mathrm{exp}} \pm 0.064_{\mathrm{BR}} \pm 
                     0.089_{\mathrm{theo}} ) \ {\mathrm{GeV}}$ and
$\lambda_1^{1S} = (-0.277 \pm 0.049_{\mathrm{stat}} \pm 0.017_{\mathrm{exp}} \pm 0.022_{\mathrm{BR}} \pm
                     0.094_{\mathrm{theo}}) \ {\mathrm{GeV^2}}$,
with a 77\% positive correlation.

The statistical and systematic errors in the extraction of the $D^{**}$ moments, the full moments, and the HQET 
parameters 
in the pole-mass scheme are presented in Table~\ref{tab:syst}. 
\begin{table*}[ht]%\scriptsize
\center
\begin{tabular}{|l|c|c|c|c|c|c|}
\hline
\vspace*{-8pt}
&&&&&&\\
Error                              & $\Delta m_1$ & $\Delta m_2$ & $\Delta M_1$ & $\Delta M_2$ & 
$\Delta \overline{\Lambda}$ & $\Delta \lambda_1$ \\
                                   &  (GeV$^2/c^4$)   & (GeV$^4/c^8$)    & (GeV$^2/c^4$)    & (GeV$^4/c^8$)    & 
(GeV)            & (GeV$^2$)          \\
\hline
Statistical                        & 0.16         & 0.69         & 0.038        & 0.26         & 0.078   & 0.057 \\
Total systematic                   & 0.08         & 0.22         & 0.068        & 0.13         & 0.091   & 0.082 \\
\hline
\hline
Mass resolution                    & 0.02         & 0.13         & 0.005        & 0.04         & 0.012   & 0.009 \\
\hline
Efficiency (data)                  & 0.03         & 0.13         & 0.006        & 0.05         & 0.014   & 0.011 \\
\hline 
Efficiency and acceptance (MC)     & 0.06         & 0.05         & 0.016        & 0.03         & 0.017   & 0.006 \\
\hline
Background scale                   & 0.01         & 0.03         & 0.002        & 0.01         & 0.003   & 0.002 \\
\hline
Background bias                    & 0.02         & 0.10         & 0.004        & 0.03         & 0.006   & 0.006 \\
\hline
Physics background                 & 0.01         & 0.02         & 0.002        & 0.01         & 0.004   & 0.002 \\
\hline
$D^+/D^{*+}$ branching ratios      & 0.01         & 0.02         & 0.002        & 0.01         & 0.004   & 0.002 \\
\hline
$D^+/D^{*+}$ efficiency            & 0.02         & 0.03         & 0.004        & 0.01         & 0.005   & 0.002 \\
\hline
$B$ semileptonic branching ratios  & ---          & ---          & 0.065        & 0.10         & 0.064   & 0.022 \\
\hline
$\rho_1$                           & ---          & ---          & ---          & ---          & 0.041   & 0.069 \\
\hline
$T_i   $                           & ---          & ---          & ---          & ---          & 0.032   & 0.031 \\
\hline
$\alpha_s$                         & ---          & ---          & ---          & ---          & 0.018   & 0.007 \\
\hline
$m_b,m_c$                          & ---          & ---          & ---          & ---          & 0.001   & 0.008 \\
\hline
Choice of $p_l^*$ cut              & ---          & ---          & ---          & ---          & 0.019   & 0.009 \\
\hline
\end{tabular}
\caption{Statistical and 
systematic uncertainties in the measurements of the $D^{**}$ and full moments and in the extraction of
$\overline{\Lambda}$ and $\lambda_1$ in the pole scheme.\label{tab:syst}}
\end{table*}
Statistical errors dominate the measurements of $m_1$ and $m_2$ while experimental systematic errors are 
all smaller.
The main experimental systematics
are computed from the differences in the results 
when we apply or omit
a correction for the $\sim 60$~MeV/c$^2$ mass resolution in the $K^-\pi^+\pi^0$ channel
or the $M(D^{**})$-dependent efficiency correction from data.
Similarly, we considered the difference in results obtained using
Monte Carlo efficiency and acceptance corrections calculated with
ISGW2 and Goity-Roberts matrix elements or phase space as a systematic error.
Other experimental systematics include uncertainties on the level of the prompt background (studied 
with a
fully reconstructed $B$ sample), a possible bias due to having used the same data sample to
model the background in the optimization process and to subtract the background from the data (studied by repeating
the optimization on bootstrap copies of the data),
and uncertainties on $D$ branching fractions used in the analysis.

Uncertainties in the inclusive and exclusive semileptonic $B$ branching ratios become important
when combining $m_1$ and $m_2$ with the $D^0$ and $D^{*0}$ pieces to obtain $M_1$ and $M_2$, the moments of the entire
charm mass distribution. Theoretical uncertainties become dominant in the extraction of the HQET parameters
$\Lambda$ and $\lambda_1$. The largest contribution to the theoretical systematic error
is that estimated by varying the unknown third order HQET parameters in the ranges
$\rho_1 = \frac12 (0.5~\mbox{\rm GeV})^3 \pm \frac12 (0.5~\mbox{\rm GeV})^3$, $T_i = (0.0~\mbox{\rm GeV})^3 
\pm (0.5~\mbox{\rm GeV})^3$.
Finally,
acceptance corrections have been computed for two alternative $p_l^*$ cuts, 0.5 and 0.9~GeV/$c$. The moments obtained
this way are different physical quantities and numerically different from those obtained for the default $p_l^*$ cut at
0.7~GeV/$c$.
However, if HQET describes the data, they should all lead to 
compatible values of the HQET parameters. The three sets of parameters are found to be equal within errors.
Differences between them have been considered as additional systematic
uncertainties.
\vspace{1em}

In summary, we have presented a measurement of the first two moments of the hadronic mass-squared distribution
in semileptonic $B$ decays to charm by combining our measurement of the $D^{(*)+}\pi^-$ 
mass spectrum %above the $D^{*0}$ mass 
with the known masses and branching ratios to $Dl\nu$ and $D^*l\nu$ taken from the Particle Data Group 
compilation~\cite{ref:pdg04}. These channels together are assumed to fully account for the inclusive
semileptonic decay width of $B$ mesons to charm.
The moments are then used to extract the two leading Heavy Quark Effective Theory parameters,
$\Lambda$ and $\lambda_1$, in both the pole mass and the 1S mass schemes. 
Within HQET the results are in agreement with previous 
determinations at $e^+e^-$ machines~\cite{ref:previous}, mostly at higher values of $p_l^*$,
and our precision in the moments is comparable or slightly better. The experimental techniques and systematic uncertainties
are very different from those at electron-positron colliders.
Our measurements, when combined with moments of the
lepton energy in semileptonic B decays~\cite{ref:lmoments} and photon energy in $b\to s\gamma$
transitions~\cite{ref:gmoments}
measured elsewhere,
will help pinpoint the values of the non-perturbative HQET parameters~\cite{ref:bauer2}.
This will sharpen the experimental 
determination
of $|V_{cb}|$ and allow tests of the underlying assumptions of the theoretical framework, such as quark-hadron
duality.
\section*{ACKNOWLEDGMENTS}
We thank the Fermilab staff and the technical staffs of the participating institutions 
for their vital contributions. It is a pleasure to thank Zoltan Ligeti for several 
enlightening discussions regarding the HQET predictions.
This work was supported by the U.S.
Department of Energy and National Science Foundation; 
the Italian Istituto Nazionale di Fisica Nucleare; the Ministry of Education, Culture, Sports,
Science and Technology of Japan; the Natural Sciences and Engineering Research Council of Canada; 
the National Science Council of the Republic
of China; the Swiss National Science Foundation; the A.P. Sloan Foundation; 
the Bundesministerium f\"ur Bildung und Forschung, Germany; the
Korean Science and Engineering Foundation and the Korean Research Foundation; 
the Particle Physics and Astronomy Research Council and the
Royal Society, UK; the Russian Foundation for Basic Research; 
the Comisi\'on Intermi\-nis\-terial de Ciencia y Tecnolog\'{\i}a, Spain; 
and in part by the European Community's Human Potential Programme under contract HPRN-CT-2002-00292, 
Probe for New Physics. 
%==============================================================================

%==============================================================================
\end{document}